\documentclass[preprint,showpacs,preprintnumbers,amsmath,amssymb]{revtex4}%

\usepackage{graphicx}
\usepackage{dcolumn}
\usepackage{bm}
\usepackage{amsmath}
\usepackage{amsfonts}
\usepackage{amssymb}%
\providecommand{\U}[1]{\protect\rule{.1in}{.1in}}

\newcommand{\be}{\begin{equation}}
\newcommand{\en}{\end{equation}}
\newcommand{\bea}{\begin{eqnarray}}
\newcommand{\ena}{\end{eqnarray}}

\begin{document}

\title{Dynamics of cosmological inflation and predictions for reheating in the light of 2018 PLANCK results}

\author{Mat\'ias L\'opez}
\email{matias.lopez.g@mail.pucv.cl}

\author{Nelson Videla}
\email{nelson.videla@pucv.cl} 

\affiliation{ Instituto de
F\'{\i}sica, Pontificia Universidad Cat\'{o}lica de Valpara\'{\i}so,
Avenida Brasil 2950, Casilla 4059, Valpara\'{\i}so, Chile.}

\author{Grigoris Panotopoulos}
\email{grigorios.panotopoulos@tecnico.ulisboa.pt} \affiliation{ Centro de Astrof\'{\i}sica e Gravita{\c c}{\~a}o, Departamento de F{\'i}sica, Instituto Superior T\'ecnico-IST, Universidade de Lisboa-UL, Av. Rovisco Pais, 1049-001 Lisboa, Portugal}

\date{\today}

\begin{abstract}
We study the dynamics of two concrete inflationary models, namely Spontaneous Symmetry Breaking Inflation  as well as Loop Inflation. We constrain the parameters for which a viable inflationary Universe emerges using the latest PLANCK results of last year, and we give predictions for the duration of reheating as well as for the reheating temperature after inflation. Our numerical results show that baryogenesis via leptogenesis may be realized within the inflationary models considered in this work.

\end{abstract}


\maketitle

\section{Introduction}

Standard Hot big-bang Cosmology, based on Einstein's General Relativity \cite{GR} and the cosmological principle, is supported by the three main pillars of Modern Cosmology, namely, the Hubble's law \cite{hubble}, the Primordial big-bang Nucleosynthesis (BBN) \cite{fowler} as well as the Cosmic Microwave Background (CMB) Radiation \cite{cmb}. Despite its success, however, it suffers from some long standing puzzles, such as the horizon, flatness, and monopole problems. Those problems find a natural explanation in the framework of cosmological inflation \cite{starobinsky1,inflation1,inflation2,inflation3}. What is more, since quantum fluctuations during the inflationary era may give rise to the primordial density perturbations with an approximately scalar-invariant power spectrum \cite{Starobinsky:1979ty,R2,R202,R203,R204,R205,Abazajian:2013vfg}, inflation provides us with a causal interpretation of the origin of the CMB temperature anisotropies, while at the same time it comes with a mechanism to explain the Large-Scale Structure (LSS) of the Universe. Therefore, currently cosmological inflation is widely accepted as the standard paradigm of the very early Universe, although we do not have a theory of inflation yet. For a classification of all single field inflationary models based on a minimally coupled scalar field see \cite{kolb}, while for a large collection of inflationary models and their connection to Particle Physics see \cite{riotto,martin}.

When the slow-roll approximation breaks down inflation ends and the Universe enters into the radiation era of standard Hot big-bang Cosmology \cite{Lyth:2009zz}. The transition era after the end of inflation, during which the inflaton is converted into the particles that populate the Universe later on is called reheating \cite{reh1,reh2}. Unfortunately the physics of reheating is complicated, highly uncertain, and in addition it cannot be directly probed by observations, although some bounds from BBN \cite{bound1,bound2}, the gravitino problem \cite{gravitino1,gravitino2,gravitino3,gravitino4} and leptogenesis \cite{leptogenesis} do exist. One may obtain, however, indirect constraints on reheating according to the following strategy: First we parameterize our ignorance assuming for the fluid a constant equation-of-state $w_{re}$ during reheating, and then we find certain relations between the reheating temperature and the duration of reheating with $w_{re}$ and inflationary observables \cite{paper1,paper2,paper3}.

In the present work we propose to study the dynamics of two concrete inflationary models, and to give predictions for the duration of reheating as well as the reheating temperature after inflation using the latest results of the PLANCK collaboration of last year. The plan of our work is the following: In the next section we briefly review the dynamics of any canonical single field inflationary model in the slow-roll approximation. In section 3 we present and discuss our main numerical results. Implications for baryogenesis are also briefly discussed. Finally we finish our work with some concluding remarks in the fourth section.

\section{Theoretical Framework}

Our starting point is the action for a single scalar field minimally coupled to gravity:
\be
S = \int d^4x \: \sqrt{-g} \left( \frac{M_{pl}^2}{2}R+K(\phi, X) \right)
\,.\label{accion}
\en
where $g$ is the determinant of metric tensor $g_{\mu\nu}$, $M_{pl}=2.4 \times 10^{18}~GeV$ is the reduced Planck mass, $R$ is the Ricci scalar, and $K(\phi,X)$ is a generic arbitrary function of both $X$ and $\phi$, with $X=g^{\mu\nu}\partial_{\mu}\phi\partial_{\nu}\phi/2$
being the standard kinetic term, and $\phi$ being the scalar field.

In the discussion to follow we shall assume i) an action with $K = X-V(\phi)$, with $V(\phi)$ being the effective potential for the scalar field, ii) a spatially flat Friedmann-Lema\^{i}tre-Robertson-Walker (FLRW) metric
\begin{equation}
ds^2 = -dt^2 + a(t)^2 \delta_{ij} dx^i dx^j
\end{equation}
with $t$ being the cosmic time, $a(t)$ being the scale factor, and where the spatial indices
$i,j=1,2,3$, and finally iii) a homogeneous scalar field $\phi=\phi(t)$. Since a homogeneous scalar field with a self-interaction potential $V(\phi)$ behaves like a perfect fluid with energy density $\rho_\phi$ and pressure $p_\phi$ \cite{riotto}
\begin{eqnarray}
\rho_\phi & = & \frac{\dot{\phi}^2}{2} + V \\
p_\phi & = & \frac{\dot{\phi}^2}{2} - V
\end{eqnarray}
respectively, the first Friedmann and the Klein-Gordon (KG) equations read
\begin{equation}
H^2 = \frac{1}{3M_{pl}^2} \left( \frac{\dot{\phi}^2}{2}+V(\phi) \right),\label{Friedmann}
\end{equation}
and
\begin{equation}
\ddot{\phi} + 3 H\dot{\phi} + V_{,\phi} = 0, \label{KG}
\end{equation}
respectively, where $,\phi$ denotes differentiation with respect to the scalar field, and $H \equiv \dot{a}/a$ is the Hubble parameter. The condition for inflation (i.e. accelerating expansion), $\ddot{a} > 0$, requires that $\dot{\phi}^{2} < V(\phi)$, or in other words that the potential energy of the inflaton field dominates over its kinetic energy. Imposing the so-called \emph{slow-roll} conditions, $\dot{\phi}^2/2 \ll V(\phi)$ and $|\ddot{\phi}| \ll 3H|\dot{\phi}|$, the Friedmann and the KG equations take the approximate form
\begin{equation}
H^2 \simeq \frac{V(\phi)}{3M_{pl}^2}, \label{FriedmannSR}
\end{equation}
and
\begin{equation}
3 H \dot{\phi} \simeq -V_{,\phi}. \label{KGSR}
\end{equation}
respectively.
One can define the so-called \emph{slow-roll} parameters as follows \cite{kolb}
\begin{equation}
\label{SRP}
\epsilon = \frac{M_{pl}^{2}}{2}\left(\frac{V_{,\phi}}{V}\right)^{2} \, , \, \, \eta=M_{pl}\,\left(\frac{V_{,\phi \phi}}{V}\right).
\end{equation}
The first slow-roll parameter is always positive by definition, while the second one may be either positive or negative depending on the shape of the scalar potential. The \emph{slow-roll} approximation requires that $\epsilon \ll 1$ and $|\eta|\ll 1$ for a prolonged amount of time. The inflationary epoch ends when $\epsilon$ or $|\eta|$ becomes unity. 

A useful quantity to describe the amount of inflation is the number
of $e$-folds, defined by \cite{kolb}
\begin{equation}
\label{Nk}
N_{k}\equiv \ln \frac{a_{end}}{a_{k}}=\int_{t_{k}}^{t_{end}} \, Hdt \simeq \frac{1}{M_{pl}^2}\,\int_{\phi_{end}}^{\phi_{k}} \, \frac{V}{V_{,\phi}} d \phi,
\end{equation}
where the sub-index $end$ denotes the end of the inflationary epoch, while the sub-index 
$k$ denotes the instant when the cosmological scale crosses the Hubble radius.

The connection between inflationary predictions and observations is made using the amplitude $A$ and the spectral index $n$ of scalar (S) and tensor (T) perturbations, which follow a power-law with the scale $k$ as follows
\begin{eqnarray}
A_S (k) & \propto & k^{n_s-1} \\
A_T(k) & \propto & k^{n_T}
\end{eqnarray}
In particular, within the \emph{slow-roll} approximation, the power spectrum of the scalar perturbation and its scalar spectral index, in terms of the \emph{slow-roll} parameters, are given by \cite{kolb}
\begin{equation}
\label{Ps}
P_{S} = \frac{V(\phi)}{24\,\pi^{2}\,\epsilon}, 
\end{equation}
\begin{equation}
\label{ns}
n_{s} = 1+2\eta-6\epsilon,
\end{equation}
respectively. In addition, another observable of paramount importance in the cosmological inflation is the tensor-to-scalar ratio $r$. It is defined as the ratio of the amplitude of tensor perturbations over the amplitude of scalar perturbations. The tensor perturbations produce the B-modes of polarization of the CMB. In this way, $r$ is given by \cite{kolb}
\begin{equation}
r = \frac{P_{T}}{P_{S}} \simeq 16 \,\epsilon, \label{r}
\end{equation}
where in the last equality the \emph{slow-roll} approximation has been used. Finally, the tensor spectral index is given by $n_T = -2 \epsilon$, and therefore it is related to $r$ through the so-called consistency relation \cite{kolb}
\begin{equation}
r = - 8 n_T
\end{equation}
and so it is not an independent parameter.

The reheating process after the inflationary epoch is important for itself 
as a mechanism to achieve what we know as the hot big-bang Universe. The energy of the inflaton field becomes in thermal radiation during the process of reheating through particle creation while the inflaton field oscillates about the minimum of its potential. What is more, the duration of the reheating $N_{re}$ as well as the reheating temperature $T_{re}$ are given by \cite{paper2,paper3}
\begin{eqnarray}
\label{Nre}
N_{re} &=& \frac{4}{1-3 w_{re}} \left[61.6 - \ln \left( \frac{V_{end}^{1/4}}{H_k} \right) - N_k \right],\\
\label{Tre}
T_{re} &=& exp\left[-\frac{3}{4} (1+w_{re}) N_{re} \right] \left( \frac{3}{10 \pi^2} \right)^{1/4} (1+g)^{1/4} V_{end}^{1/4},
\end{eqnarray}
with $g \approx 0.5$. Here, the model-dependent expressions are the Hubble rate at the instant when the cosmological scale crosses the Hubble radius, $H_k=\pi\sqrt{8\,P_S\,\epsilon_k}$, and the inflaton potential at the end of the inflationary expansion, $V_{end}$. For more details and on the derivation of the expressions the interested reader may consult \cite{paper2,paper3}.

Thus, it is implicit that $N_{re}, T_{re}$ depend on the observables $P_{s}$, $n_{s}$ and $r$ that we have already discussed.

\section{Numerical Results}

Our principal goal here is to study the compatibility of two concrete models with the observational constraints on the amplitude of the scalar perturbations, its scalar spectral index, and the tensor-to-scalar ratio, which values are constrained by current observations by the PLANCK collaboration \cite{planck2,planck4} as well as the BICEP2/Keck-Array data \cite{bicep,Ade:2018gkx}.

Any inflationary model based on a single canonical scalar field is defined by assuming a scalar potential for the inflaton. In the present work we shall consider i) Spontaneous Symmetry Breaking Inflation (or Higgs inflation) defined by the potential \cite{martin}
\begin{equation}
V(\phi)= V_{0} + A \phi^2 + B \phi^4
\end{equation}
which may be viewed as a Taylor expansion of a generic scalar potential $V(\phi)$, and where the parameters $A,B$ a priori can be either positive or negative, and ii) Loop Inflation defined by the potential \cite{martin}
\begin{equation}
V(\phi)= M^4 \: \left[1 + a \textrm{ln} \left( \frac{\phi}{M_{pl}} \right)  \right]\label{LI}
\end{equation}
characterized by a mass scale $M$ and a dimensionless parameter $a$. The shape of the scalar potential Loop Inflation may arise for instance in supersymmetric models \cite{wess,primer}, and in particular in D-term inflation \cite{dterm1,dterm2,dterm3}. 

The polynomial potential for the inflaton has been analysed by several authors, see e.g. \cite{SSBI1,SSBI2,SSBI3,SSBI4}. In particular, in the following we shall consider the case where the mass term enters with the wrong sign \cite{Kobayashi:2014jga}
\begin{equation}
V(\phi)= V_{0} - \frac{1}{2}\,m^{2}\,\phi^{2} + \frac{\lambda}{4}\,\phi^{4},
\end{equation}
characterized by a mass scale $m$ and a dimensionless coupling $\lambda$. From the condition of a vanishing cosmological constant at the minimum, it is found that the vacuum expectation value (VEV) of $\phi$ and $V_{0}$ become
\begin{equation}
\langle \phi \rangle^2 = \frac{m^{2}}{\lambda} \,; V_{0}=\frac{m^{4}}{4\,\lambda}.
\end{equation}
Studying the behaviour of \emph{slow-roll} parameters, we notice that the inflationary epoch ends when $\eta$ becomes unity, since this quantity evolves faster than $\epsilon$. From Eq.(\ref{SRP}), the inflaton field has the following value when inflation comes to an end
\begin{equation}
\left(\frac{\phi_{end}}{M_{pl}}\right)^{2}=6+\alpha-2\sqrt{9+2\alpha},\label{phiend}
\end{equation}
where $\alpha$ is a dimensionless parameter defined to be
\begin{equation}
\alpha = \frac{m^2}{\lambda M_{pl}^2}.\label{alpha}
\end{equation}
Solving Eq.(\ref{Nk}) one obtains the value of the inflaton field for any cosmological scale $k$ as a function of the number of $e$-folds $N_{k}$ and $\alpha$, yielding
\begin{equation}
y_k\equiv \frac{\phi_k}{M_{pl}}=\alpha^{1/2}\,W^{1/2}[z(N_k,\alpha)],\label{yk}
\end{equation}
where
\begin{equation}
z(N_k,\alpha)=\frac{1}{\alpha}e^{-\frac{1}{\alpha}\left(8N_k+6+\alpha-2\sqrt{9+2\alpha}\right)}\left(8N_k+6+\alpha-2\sqrt{9+2\alpha}\right)\label{zNk}
\end{equation}
with $W(z)$ being the Lambert function \cite{Veberic:2012ax}

Then, after plugging Eqs.(\ref{yk}) into Eqs.(\ref{Ps}), (\ref{ns}), and (\ref{r}), the
inflationary observables $P_{S}$, $n_{s}$ and $r$ are computed to be
\begin{eqnarray}
P_{S} & = & \frac{\lambda}{768 \pi^2 y_k^2}\left(\alpha-y_k^2\right)^4,\label{PSk} \\
n_s & = & 1-8\frac{(\alpha+3y_k)}{\left(\alpha-y_k^2\right)^2},\label{nsk} \\
r & = & 128 \frac{y_k^2}{\left(\alpha-y_k^2\right)^2}\label{rk},
\end{eqnarray}

\begin{figure}[ht!]
\centering
\includegraphics[scale=0.7]{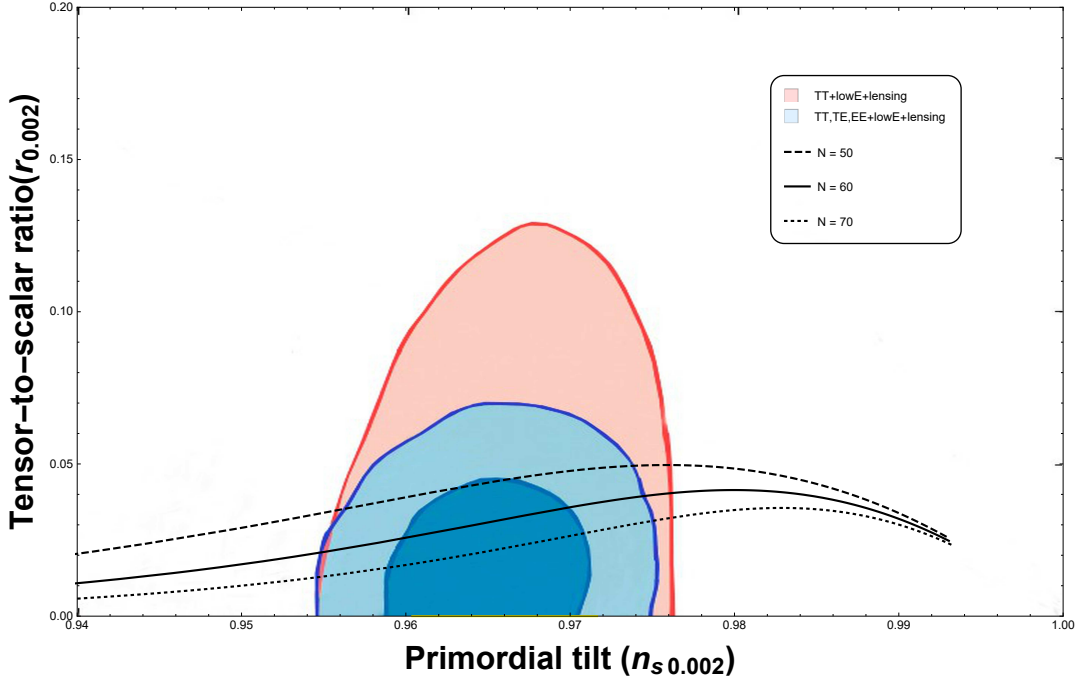}
{\vspace{0 in}}
\caption{Allowed contours at the 68 and 95 $\%$ C.L., from the latest PLANCK data \cite{planck4}, and the theoretical predictions in the $n_{s}-r$ plane for the polynomial inflationary model. We have used three different values for the number of $e$-folds $N_k$: $N_k=50$ (dashed), $N_k=60$ (solid), and $N_k=70$ (dotted).}
\label{fig1} 	
\end{figure}

Next, the trajectories in the $n_s-r$ plane for the models studied here may be generated by plotting Eqs.(\ref{nsk}) and (\ref{rk}) parametrically allowing a wide range for $\alpha,a$. In particular, we have obtained three different curves setting the number of $e$-folds to be $N_k=50,60,70$. In Fig.~\ref{fig1} and \ref{fig3} we show the plot of the tensor-to-scalar ratio $r$ versus the scalar spectral index $n_s$ for the polynomial inflation and for Loop Inflation, respectively. Here, we have considered the two-dimensional (2D) marginalized joint confidence contours for ($n_s,r$), at the 68 $\%$ and 95 $\%$ CL, from the latest PLANCK data of last year \cite{planck4}. The corresponding allowed range for the dimensionless parameters $a$ and $\alpha$ defined by Eq.(\ref{alpha}), for each $r(n_s)$ curve, may be inferred by finding the points at which the trajectory enters and exits the 95 $\%$ CL region from the PLANCK data. 

Regarding the polynomial potential first, making use of the Eqs.(\ref{nsk}) and (\ref{rk}), we have checked numerically that, for a given $N_k$, the scalar spectral index increases while $\alpha$ increases. On the other hand, after reaching a maximum value, the tensor-to-scalar ratio reaches a maximum value and after that it starts to decrease as $\alpha$ increases. That way, we may obtain both a lower and an upper bound for the $\alpha$ parameter. For each case, as $N_k$ increases the shown curves lead to lower tensor-to-scalar ratio. The theoretical predictions lie inside the allowed region from PLANCK when $\alpha$ takes values in the following range:
\\
For $N_k=50$,
\begin{equation}
2.15 \times 10^2 < \alpha < 4.20 \times 10^2.\label{range_alpha1}
\end{equation}
For $N_k=60$,
\begin{equation}
1.95 \times 10^2 < \alpha < 2.15 \times 10^2.\label{range_alpha2}
\end{equation}
For $N_k=70$,
\begin{equation}
1.90 \times 10^2 < \alpha < 4.00 \times 10^2.\label{range_alpha3}
\end{equation}
Furthermore, by combining the scalar power spectrum (\ref{PSk}), the constraint on $\alpha$ already obtained, and the observational value of the amplitude of the scalar power spectrum $P_{S} \simeq 2.15 \times 10^{-9}$, we may infer the allowed range for the mass scale $m$ and the coupling $\lambda$ for each value of $N_k$.
\\
For $N_k=50$,
\begin{equation}
4.84 \times 10^{-6} \, M_{pl} > m > 4.42 \times 10^{-6} \, M_{pl},\label{range_m1}
\end{equation}
and accordingly
\begin{equation}
1.09 \times 10^{-13} > \lambda > 4.65 \times 10^{-14}.\label{range_lambda1}
\end{equation}
For $N_k=60$,
\begin{equation}
3.81 \times 10^{-6} \,M_{pl} < m < 3.90 \times 10^{-6} \, M_{pl},\label{range_m2}
\end{equation}
and accordingly
\begin{equation}
7.44 \times 10^{-14} > \lambda > 3.71 \times 10^{-14}.\label{range_lambda2}
\end{equation}
For $N_k=70$,
\begin{equation}
3.01 \times 10^{-6} \, M_{pl} < m < 3.45 \times 10^{-6} \, M_{pl},\label{range_m3}
\end{equation}
and accordingly
\begin{equation}
4.77 \times 10^{-14} > \lambda > 2.98 \times 10^{-14}.\label{range_lambda3}
\end{equation}

Regarding Loop Inflation potential (\ref{LI}), we notice that inflation ends when $\epsilon$
becomes unity before than $\eta$. From Eq.(\ref{SRP}), the inflaton field has the following value
when inflation comes to a end
\begin{equation}
\label{LIend}
\phi_{end}=\frac{1}{\sqrt{2}}\left(W\left[\frac{e^{1/a}}{\sqrt{2}}\right]\right)^{-1},
\end{equation}
with $W$ denoting the Lambert function.

The value of the inflaton field
for any cosmological scale $k$, as a function of $a$ and the number of $e$-folds $N_k$, 
is obtained upon solving the following equation for $y_k=\frac{\phi_k}{M_{pl}}$ which arise by combining Eqs.(\ref{Nk}), (\ref{LI}), and (\ref{LIend})
\begin{equation}
4aN_k=y_k^2\left(2-a+2a\ln y_k\right)-y_{end}^2\left(2-a+2a\ln y_{end}\right),\label{ykli}
\end{equation}
where $y_{end}=\frac{\phi_{end}}{M_{pl}}$.

Then, after plugging the solution for $y_k$ into Eqs.(\ref{Ps}), (\ref{ns}), and (\ref{r}), the
inflationary observables $P_{S}$, $n_{s}$ and $r$ become as follows
\begin{eqnarray}
P_S&=&\frac{y_k^2\,M^4}{12\,a^2\,\pi^2}\left(1+a\ln y_k\right)^3,\label{Pskli}\\
n_s&=&1-a\frac{\left(2+3a+2a\,\ln y_k\right)}{\left(y_k+ay_k\,\ln y_k\right)^2},\label{nskli}\\
r&=&\frac{8 a^2}{\left(y_k+ay_k\,\ln y_k\right)^2}.\label{rkli}
\end{eqnarray}

According to Fig.~\ref{fig3} the $N_k=60,70$ curves lie outside the allowed region and therefore they must be ruled out. For $N_k=50$, following a similar analysis, we obtain for $a$ the lower bound $a > 0.8$, and for $M$ the upper bound $(M/M_{pl}) < 0.004$.
\begin{equation}
a > 0.8 \label{range_a}
\end{equation}
\begin{equation}
M < 4 \times 10^{-3} \: M_{pl},\label{range_M}
\end{equation}
where the latter comes from Eq.(\ref{Pskli}).


\begin{figure}[ht!]
\centering
\includegraphics[scale=0.85]{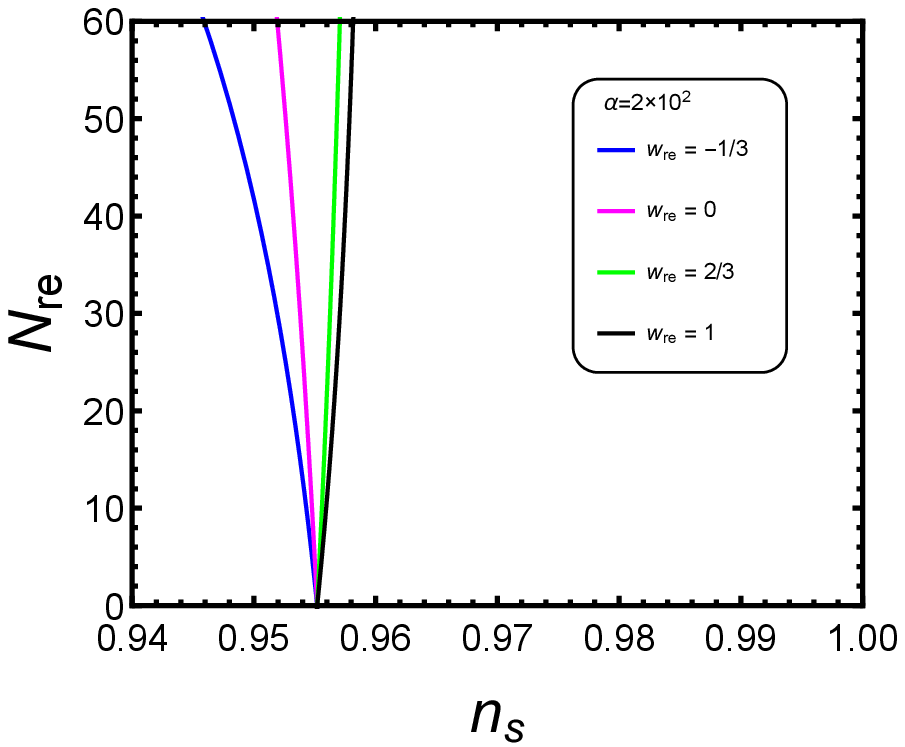}
\includegraphics[scale=0.87]{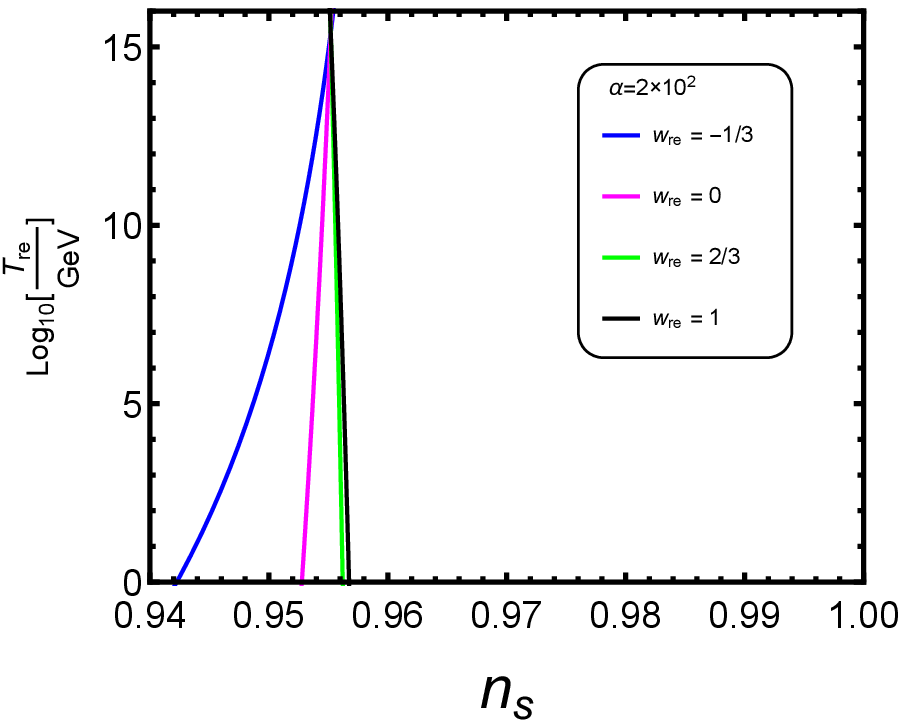}
\includegraphics[scale=0.85]{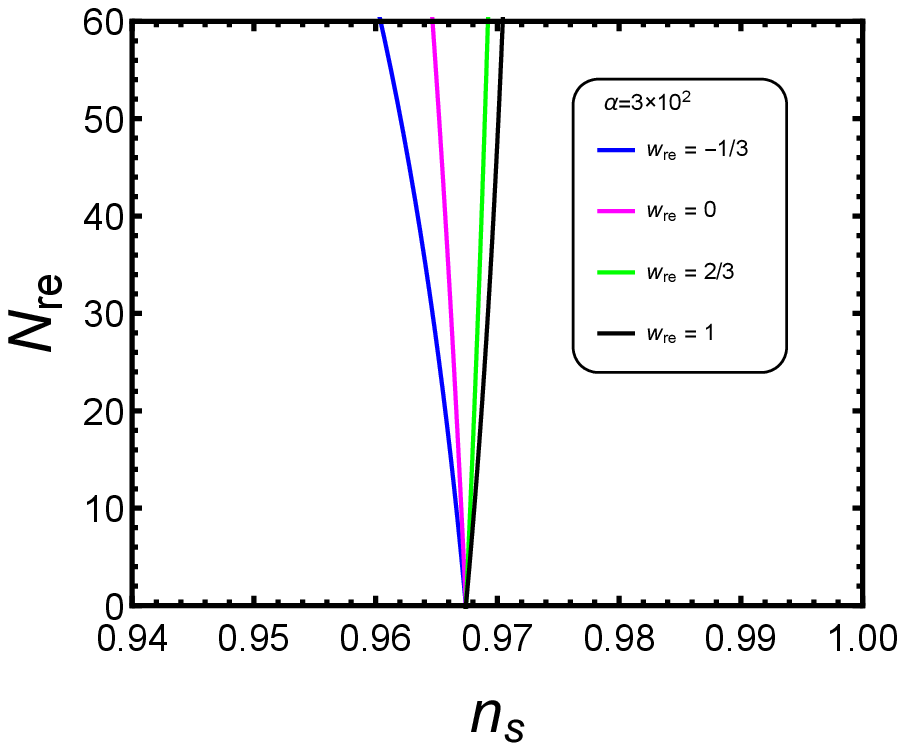}
\includegraphics[scale=0.87]{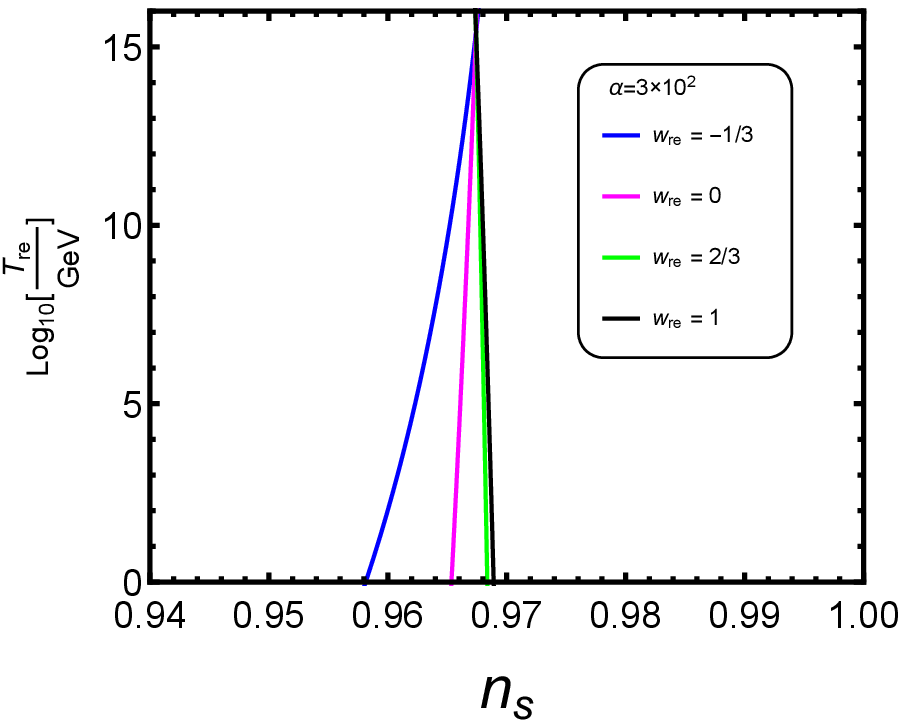}
\includegraphics[scale=0.85]{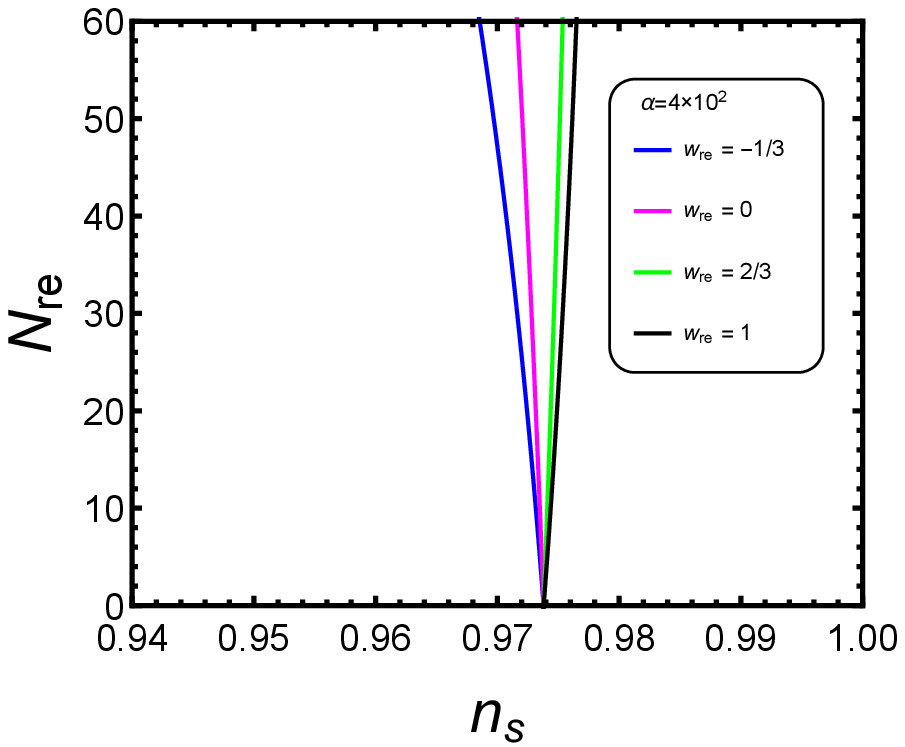}
\includegraphics[scale=0.87]{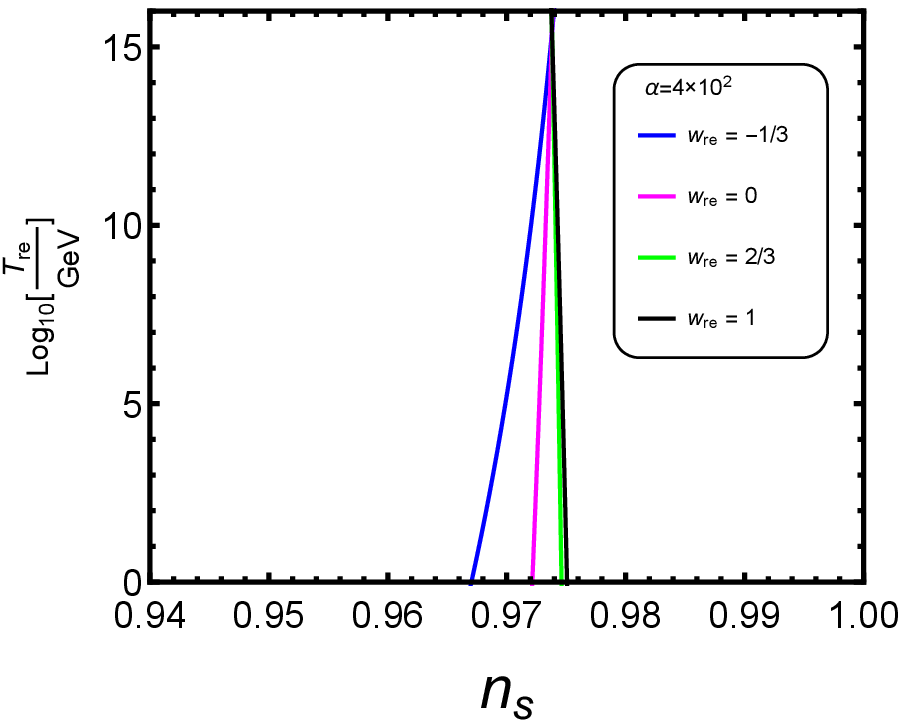}
{\vspace{0 in}}
\caption{Plots  of $N_{re}$ and $T_{re}$, the duration of reheating (left column)
and the temperature (right column) at the end of reheating respectively, for the polynomial potential.}
\label{fig2} 	
\end{figure}

\begin{figure}[ht!]
\centering
\includegraphics[scale=0.7]{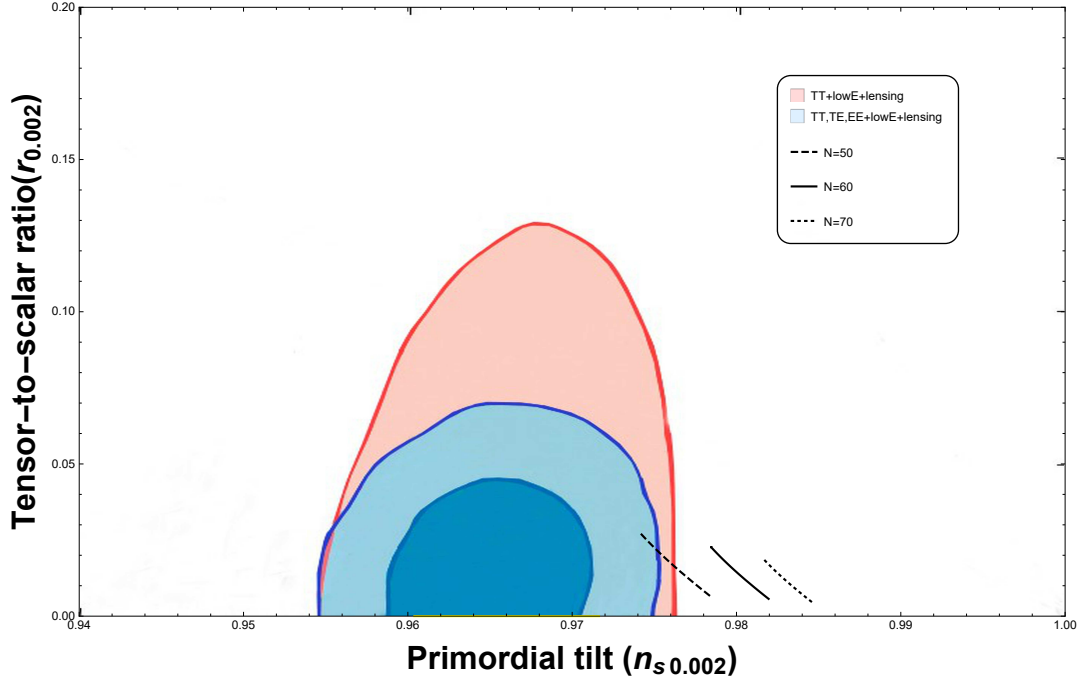}
{\vspace{0 in}}
\caption{Same as in Fig.~1, but for Loop Inflation. Only the $N_k=50$ 
curve lies inside the allowed region.}
\label{fig3} 	
\end{figure}

\begin{figure}[ht!]
\centering
\includegraphics[scale=0.85]{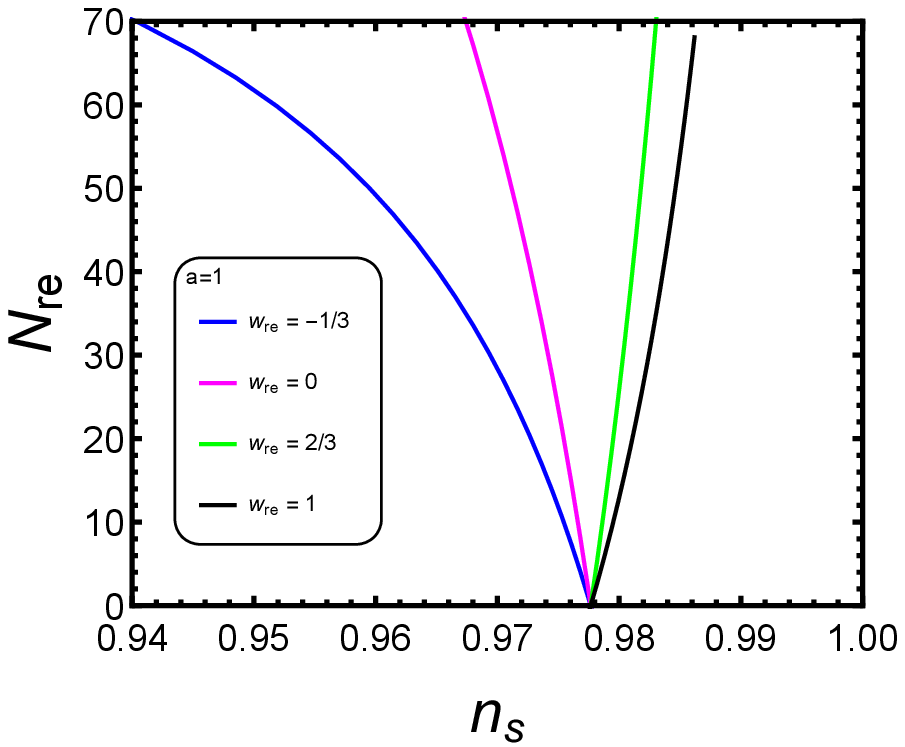}
\includegraphics[scale=0.87]{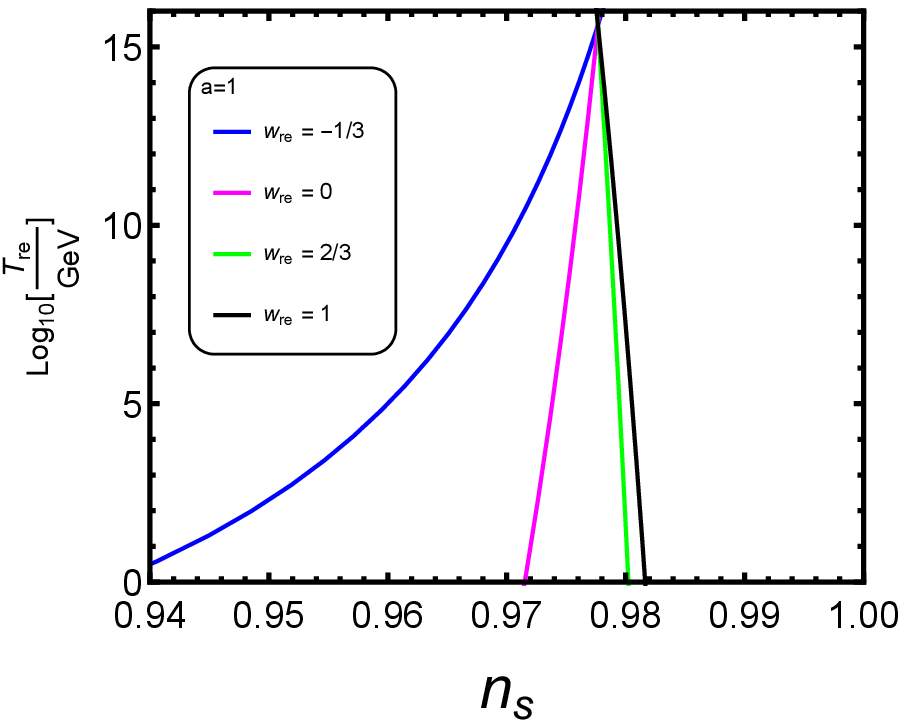}
\includegraphics[scale=0.85]{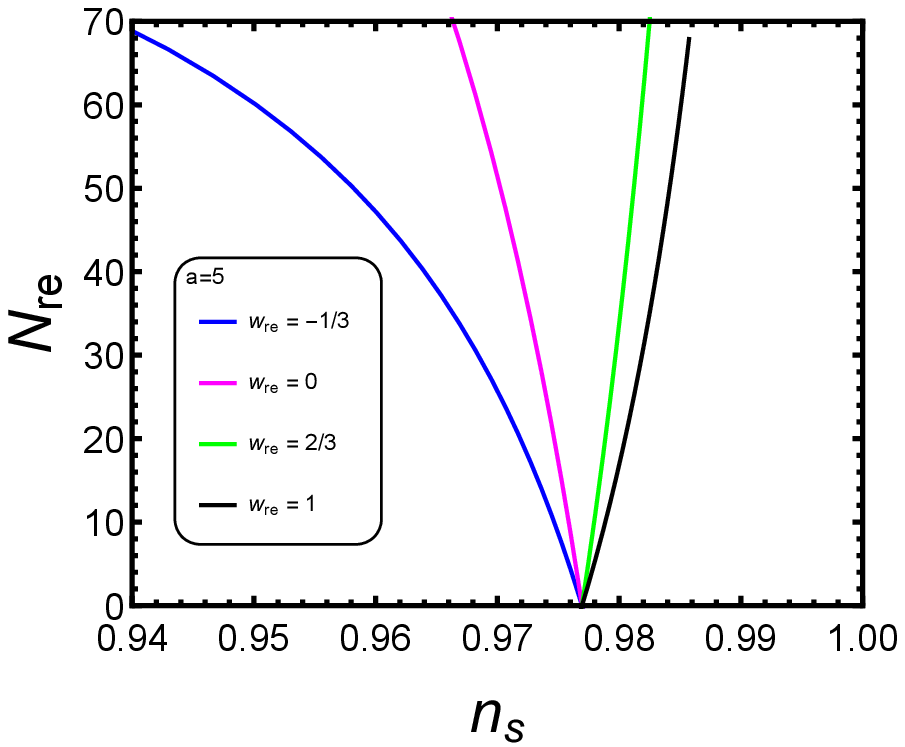}
\includegraphics[scale=0.87]{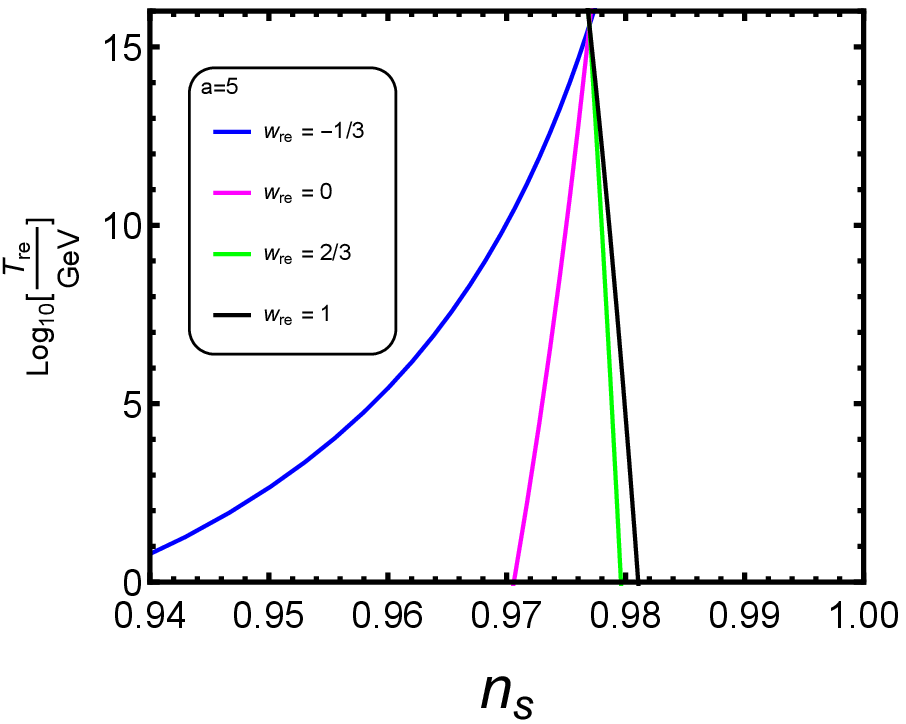}
\includegraphics[scale=0.85]{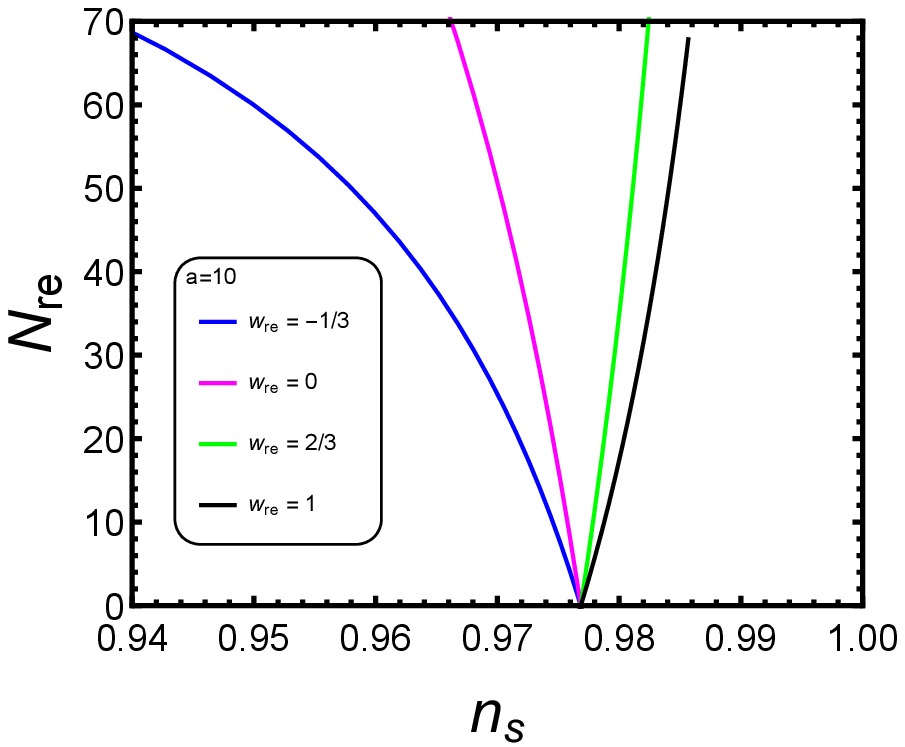}
\includegraphics[scale=0.87]{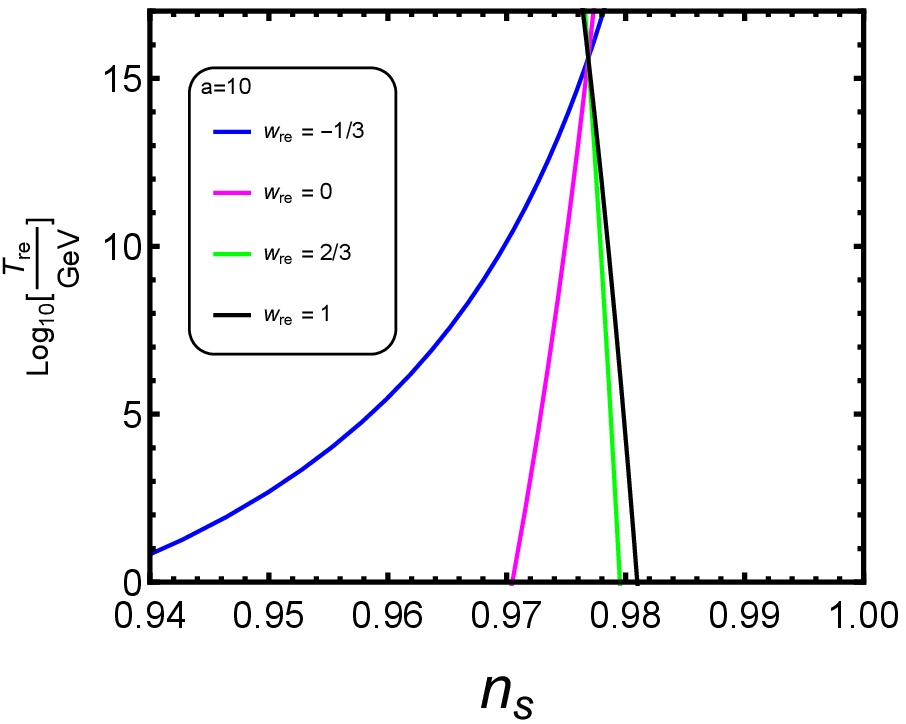}
{\vspace{0 in}}
\caption{Same as in Fig.~2, but for Loop Inflation.}
\label{fig4} 	
\end{figure}

Finally, we now investigate the predictions regarding the number of $e$-folds as well as the temperature associated with the reheating epoch $N_{re}$ and $T_{re}$, respectively. Our results are shown in Fig.~\ref{fig2} for the polynomial potential, and in Fig. \ref{fig4} for Loop Inflation. In doing so, we assume that during this epoch, the universe is governed by an effective equation-of-state of the form $P = w_{re} \rho$, where $P$ and $\rho$ denote the pressure and the energy density, respectively, of the fluid in which the inflaton decays. Then, we may plot $N_{re}$ and $T_{re}$ versus the scalar spectral index for several values of the effective equation-of-state parameter $w_{re}$ over the range $-\frac{1}{3}\leq w_{re}\leq1$, as well as $\alpha$ which encodes the information about the mass scales $m$ and the coupling $\lambda$. In the case of the polynomial potential, as $\alpha$ increases the curves are shifted to the right. On the contrary, in the Loop Inflation case, the predictions are very little sensitive to $a$.


Before concluding our work, let us here briefly comment on the implications for baryogenesis.
One of the goals of any successful inflationary model must be the explanation of the 
baryon asymmetry in the Universe, which comprises one of the biggest challenges of modern theoretical cosmology. Both primordial big-bang nucleosynthesis \cite{bbn} and data from the CMB temperature anisotropies \cite{planck1,planck3,wmap} show that the baryon-to-photon ratio is a tiny number, $\eta_B=6.19 \times 10^{-10}$ \cite{values}. This number should be calculable within the framework of known particle physics. Although several mechanisms exist, perhaps the most elegant one is leptogenesis \cite{leptogenesis}. A lepton asymmetry via the out-of-equilibrium decays of right-handed neutrinos is generated first, and then this lepton asymmetry is partially converted into baryon asymmetry via non-perturbative "sphaleron" effects \cite{sphalerons}. Thermal leptogenesis requires a high reheating temperature after inflation, $T_{re} > 10^9~GeV$ \cite{buchmuller,davidson}, while non-thermal leptogenesis \cite{lepto1,lepto2,lepto3,lepto4,Panotopoulos:2018pxw,values} can be realized at lower reheating temperature, $T_{re}=(10^6-10^7)~GeV$. Our numerical results for $T_{re}$ shown in the right panels of Figs.~\ref{fig2} and \ref{fig4} indicate that both leptogenesis mechanisms may be realized within the framework of the inflationary models studied in the present work.

\section{Conclusions}

We have studied the dynamics of two inflationary models based on a single canonical scalar field in the slow-roll approximation. In particular, we have considered i) a polynomial potential for the inflaton (Spontaneous Symmetry Breaking Inflation) and a logarithmic term plus a constant (Loop Inflation) found in \cite{martin}. Both models are characterized by two free parameters, namely a mass scale $m$ or $M$ and a dimensionless parameter/coupling constant $\lambda$ or $a$. First, using the latest PLANCK data we obtained the allowed range for both free parameters of the model. After that, we computed the reheating temperature $T_{re}$ as well as the duration of reheating $N_{re}$ versus the scalar spectral index $n_s$ assuming four different values for the equation-of-state parameter $w_{re}=-1/3, 0, 2/3, 1$ of the fluid into which the inflaton decays. Our results show that the reheating temperature is sufficiently high to support both non-thermal and thermal leptogenesis.


\section*{Acknowlegements}

The author N.~V. was supported by Comisi\'on Nacional de Ciencias y Tecnolog\'ia of Chile through FONDECYT Grant N$^{\textup{o}}$ 11170162. He also thanks the Instituto Superior T\'ecnico of Universidade de Lisboa, where part of the work was completed, for its warmest hospitality.
The author G.~P. thanks the Funda\c c\~ao para a Ci\^encia e Tecnologia (FCT), Portugal, for the financial support to the Center for Astrophysics and Gravitation-CENTRA, Instituto Superior T{\'e}cnico,  Universidade de Lisboa, through the Grant No. UID/FIS/00099/2013.


\end{document}